\documentclass[aps,twocolumn,prd,superscriptaddress,preprintnumbers,showpacs]{revtex4}
\usepackage{graphicx}

\newcommand{\be}{\begin{equation}}
\newcommand{\ee}{\end{equation}}
\newcommand{\bea}{\begin{eqnarray}}
\newcommand{\eea}{\end{eqnarray}}

\input{tcilatex}

\begin{document}

\preprint{}
\preprint{}
\title{Dynamics of Logamediate Inflation}
\author{John D. Barrow}
\email[]{J.D.Barrow@damtp.cam.ac.uk}
\affiliation{Department of Applied Mathematics and Theoretical Physics, Wilberforce Road,
University of Cambridge, Cambridge CB3 0WA, United Kingdom}
\author{N. J. Nunes}
\email[]{nunes@damtp.cam.ac.uk}
\affiliation{Department of Applied Mathematics and Theoretical Physics, Wilberforce Road,
University of Cambridge, Cambridge CB3 0WA, United Kingdom}
\date{\today}

\begin{abstract}
A computation of the inflationary observables $n_{s}$ and $r$ is made for
`logamediate' inflation where the cosmological scale factor expands as
$a=\exp (A(\ln t)^{\lambda }),$ and is compared to their predicted values in
the intermediate inflationary theory, where $a=\exp (Bt^{f})$. Both versions
prove to be consistent with observational measurements of the cosmic
background radiation. It is shown that the dynamics of a single inflaton
field can be mimicked by a system of several fields in an analogous manner
to that created by the joint evolution of the fields in assisted power-law
inflation.
\end{abstract}

\maketitle



\section{Introduction}

In the light of the latest observations of the cosmic microwave background
(CMB) radiation \cite{wmap3}, a scenario of inflationary expansion in the
early universe stands out a strong candidate to solve the horizon and
flatness problems of the standard model of cosmology as well as providing
the seeds for the formation of large-scale structure with a spectrum of
adiabatic, and nearly scale-invariant Gaussian density perturbations. For a
review of models of inflation see e.g. Ref.~ \cite{Liddle:2000cg}.

In the particular scenario of `intermediate' inflation \cite{Barrow:1990vx,saich,Muslimov:1990be,Barrow:1993zq,Vallinotto:2003vf,Rendall:2005if,Barrow:2006dh},
the expansion scale factor of the Friedmann universe evolves as $a=\exp
(At^{f})$, where $A>0$ and $1>f>0$ are constants; the expansion of the
Universe is slower for standard de Sitter, which arises when $f=1$, but
faster than in power-law inflation, $a=t^{p},$ with $p>1$ constant. It has
been shown that intermediate inflation satisfies the bounds on the spectral
index $n_{s}$ and ratio of tensor to scalar perturbations, $r$, as measured
by the latest observations of the CMB \cite{wmap3}. To first order, an exact
Harrison-Zeldovich spectrum \cite{zeld} of fluctuations arise when $f=2/3$
as well as when $f=1$. For the construction by series of a potential which
produces this spectrum to all orders, see Ref.~\cite{starob}.

In this work, we analyze another generalized version of inflation, which we
call `logamediate inflation', with scale factor of the form $a=\exp (A(\ln
t)^{\lambda })$, with $A>0$, $\lambda >1$ constants.
When $\lambda =1,$ this model reduces to
power-law inflation with $a=t^{p}$, where $p=A$, here.
The logamediate inflationary form is motivated by considering a
class of possible cosmological solutions with indefinite expansion 
which result from imposing  weak general conditions on the cosmological model. 
In Ref.~\cite{Barrow:1996bd} it was shown that an
application of the considerations of Hardy and Fowler to ordinary
differential equations of the form $\ddot{a}=P(a,t)/Q(a,t)$, as
$t\rightarrow \infty $ with polynomials $P$ and $Q$,
leads to eight possible asymptotic solutions of the cosmological dynamics.
Three of these give non-inflationary expansions for $a(t)$ and three
of the others give power-law, de Sitter, and intermediate inflationary
expansions. The remaining two cases give asymptotic expansion of the form
$a=\exp (A(\ln t)^{\lambda }),$ and this is the only one of the allowed
possibilities that, to date, has not been studied in detail with respect to the
observational data.  We note also that this form of inflation arises naturally in a number of scalar-tensor theories \cite{Barrow:1995fj}.
Here, we show that, for observationally viable
models of logamediate inflation, the ratio of tensor to scalar perturbations, $r$, must be small and
that the power spectrum can be either red or blue tilted, depending on the
specific parameters of the model.

We will also study the dynamics when an ensemble of fields is present. We
look for situations in which the dynamics allows the ratios of the
individual kinetic energies of the fields to approach constant values. This
situation has been studied in the literature in the context of power-law
inflation and it was dubbed, assisted inflation
\cite{Liddle:1998jc,Malik:1998gy,Kanti:1999vt,Copeland:1999cs,Coley:1999mj,Aguirregabiria:2000hx,Hartong:2006rt,Tsujikawa:2006mw}. It has the interesting property that the cooperative evolution of all the
fields can lead to inflation even if the individual logarithmic slopes of
the fields are too step to provide inflation when the fields are rolling in
isolation. In other words, the effective $p$ can become larger than unity if
additional fields are included in the dynamics, even if the individual $p_{i}
$ are smaller than unity if the fields were rolling alone. In the case of
logamediate inflation, as we shall see, we do not encounter this property as
the condition for inflation is set by the value of $\lambda $ alone,
nonetheless, the fractional Hubble constant, $A$, becomes the effective
quantity that is changed by the number of contributing fields.

\section{Logamediate inflation: $a=\exp \left( A(\ln t)^{\protect\lambda %
}\right) $}

We start by considering the evolution of the scale factor of a flat
Friedmann universe to be 
\begin{equation}
a(t)=\exp \left[ A\left( \ln t\right) ^{\lambda }\right] \,,  \label{afunc}
\end{equation}%
and $t>1$. The Hubble rate is: 
\begin{equation}
H\equiv \frac{\dot{a}}{a}=A\lambda \left( \ln t\right) ^{\lambda -1}\frac{1}{%
t}\,,  \label{hubble}
\end{equation}%
hence for an expanding universe we require $A\lambda >0$. On the other hand, 
\begin{eqnarray}
\frac{\ddot{a}}{a} &=&\frac{A\lambda }{t^{2}}\left( \ln t\right) ^{\lambda
-1}\left[ (\lambda -1)\left( \ln t\right) ^{-1}-1\right.  \nonumber \\
&~&\left. +A\lambda \left( \ln t\right) ^{\lambda -1}\right] \,,
\end{eqnarray}%
Therefore, for an inflationary universe, with $\ddot{a}/a>0$, it is
necessary that $\lambda >1$, or if $\lambda =1$, that $A>1$.

\subsection{Single-field inflation}

Now assume the material content of the universe is a singe scalar field $%
\phi $ with potential $V(\phi )$. Since $\dot{H}=-\dot{\phi}^{2}/2,$ at late
times we have 
\begin{equation}
\dot{\phi}=\sqrt{2}(A\lambda )^{1/2}\left( \ln t\right) ^{(\lambda -1)/2}%
\frac{1}{t}\,,  \label{phidotsol}
\end{equation}%
and the evolution of the field satisfies 
\begin{equation}
\phi =\phi _{0}+2\frac{(A\lambda )^{1/2}}{\lambda +1}\left( \ln t\right)
^{(\lambda +1)/2}\,,  \label{phisol}
\end{equation}
where $\phi _{0}$ is constant. The scalar potential results from $V=3H^{2}+%
\dot{H}$. Substituting for $H$ and $\dot{H}$ gives 
\begin{eqnarray}
V(\phi ) &=&3(A\lambda )^{2}\left( \ln t\right) ^{2(\lambda -1)}\frac{1}{%
t^{2}}+  \nonumber  \label{v_t} \\
&~&A\lambda (\lambda -1)\left( \ln t\right) ^{\lambda -2}\frac{1}{t^{2}}%
-A\lambda \left( \ln t\right) ^{\lambda -1}\frac{1}{t}\,.
\end{eqnarray}%
At late times, only the first term survives, hence, using solution (\ref%
{phidotsol}) (setting $\phi _{0}=0,$ without loss of generality), the
potential can be written as 
\begin{equation}
V(\phi )=V_{0}\phi ^{\alpha }\exp \left( -\beta \phi ^{\gamma }\right) \,,
\end{equation}%
where 
\begin{equation}
V_{0}=3(A\lambda )^{2}B^{2(\lambda -1)}\,,
\end{equation}%
and 
\begin{equation}
B=\left( \frac{\lambda +1}{2\sqrt{2}(A\lambda )^{1/2}}\right) ^{2/(\lambda
+1)}\,,
\end{equation}%
and $\alpha =4(\lambda -1)/(\lambda +1)$, $\beta =2B$, $\gamma =2/(\lambda
+1)$. This class of scalar potentials were studied in Refs.~\cite%
{Barrow:1993hn,Parsons:1995ew}. Note that we would have obtained the same
form of the scalar potential had we assumed slow-roll inflation, $3H\dot{\phi%
}\approx -dV/d\phi $. Indeed, as the field rolls down the potential towards
larger values, the slow-roll approximation becomes increasingly more
accurate, hence the two different approaches give the same result.

In the Hamilton-Jacobi formalism we write the slow-roll parameters as 
\[
\epsilon =2\left( \frac{H^{\prime }}{H}\right) ^{2}\,,\hspace{1cm}\eta =2%
\frac{H^{\prime \prime }}{H}\,, 
\]%
where prime represents differentiation with respect to the scalar field $%
\phi $. For our scalar potential these become, 
\begin{eqnarray}
\epsilon &=&\frac{1}{2\phi ^{2}}\left( \alpha -\beta \gamma \phi ^{\gamma
}\right) ^{2} \\
\eta &=&-\frac{1}{\phi ^{2}}\left[ \alpha +\beta \gamma (\gamma -1)\phi
^{\gamma }-\frac{1}{2}\left( \alpha -\beta \gamma \phi ^{\gamma }\right) ^{2}%
\right] \,.
\end{eqnarray}

The slow-roll parameter $\epsilon $ diverges when the field approaches zero,
has a minimum at the maximum of the potential, peaks at some value $\phi
_{\epsilon }$ and finally asymptotes to zero for large values of the field.
We will focus on those cases where the peak occurs for $\epsilon >1$, so
that we can identify the moment when inflation begins with $\phi _{1}\equiv
\phi (\epsilon =1)$. We are, therefore, limiting our analysis to the region
of parameter space defined by 
\begin{equation}
\beta >2\left( \frac{1}{32}(\lambda +1)^{(\lambda +3)}\right) ^{1/(\lambda
+1)}\,.
\end{equation}%
The number of e-folds between two values of the field, $\phi _{1}$ (defined
to be the beginning of inflation) and $\phi _{2}$ (when a given mode exits
the horizon) is given by 
\begin{equation}
N=-\int_{\phi _{1}}^{\phi _{2}}d\phi \,\frac{\phi }{\alpha -\beta \gamma
\phi ^{\gamma }}\,,
\end{equation}

The spectral index, $n_{s}$, and the ratio of tensor-to-scalar
perturbations, $r$, can be expressed in terms of the slow-roll parameters as 
\begin{eqnarray}
n_{s} &=&1-4\epsilon +2\eta \,, \\
r &=&16\epsilon \,.
\end{eqnarray}

\begin{figure}[tbp]
\includegraphics[width = 8.5cm]{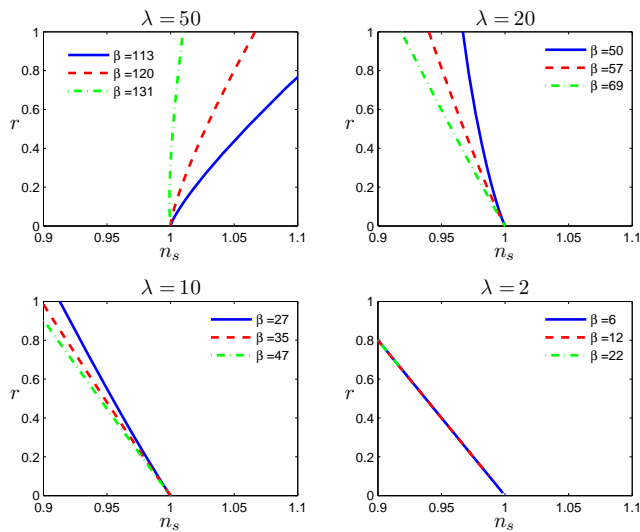}
\caption{{Trajectories for different combinations of the parameters $(%
\protect\lambda ,\protect\beta )$ in the $n_{s}-r$ plane. For these values
of $\protect\beta$, the parameter $A$ ranges form $A = 1.5 \times 10^{-92}$
for ($\protect\lambda$,$\protect\beta$) = (50,131) up to $A = 2.1 \times
10^{-2}$ for ($\protect\lambda$,$\protect\beta$) = (2,6).}}
\label{r-ns}
\end{figure}

\begin{figure}[tbp]
\includegraphics[width = 8.5cm]{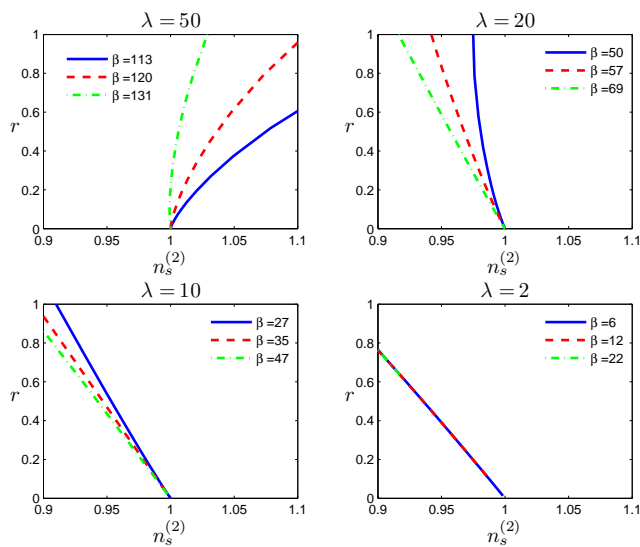}
\caption{{Trajectories for different combinations of the parameters $(%
\protect\lambda ,\protect\beta )$ in the $n_{s}^{(2)}-r$ plane, where $%
n_{s}^{(2)}$ is the second-order expansion of the spectral index in
slow-roll parameters.}}
\label{r-ns2}
\end{figure}

For a given set of parameters, $\beta $ and $\lambda $, and consequently $%
\phi _{1}$, we have fixed a value $\phi _{2}$, and then calculated the
corresponding quantities $N$, $\epsilon $, $\eta $ and $\xi ^{2}=\epsilon
\eta -(2\epsilon )^{1/2}\eta ^{\prime }$, numerically. In Fig.~\ref{r-ns} we
show the trajectories in the $n_{s}-r$ plane. Curiously, a scale-invariant
spectrum with large $r$ can be obtained for $(\lambda ,\beta )=(50,131)$. 
The second-order expression for the spectral index in terms of the slow-roll
parameters is given by 
\begin{equation}
n_{s}=1-4\epsilon +2\eta -\left[ 8(1+C)\epsilon ^{2}-(6+10C)\epsilon \eta
+2C\xi ^{2}\right] \,,
\end{equation}%
where $C=-0.73$. Comparing Figs.~\ref{r-ns} and \ref{r-ns2}, we can conclude
that the second-order correction to the spectral index is negligible. 

\begin{figure}[tbp]
\includegraphics[width = 8.5cm]{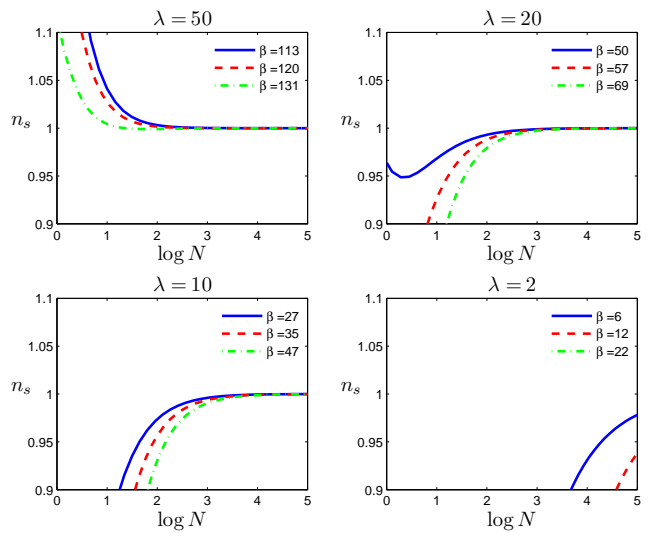}
\caption{Dependence of the spectral index on the number of e-folds of
inflation for different combinations of the parameters $(\protect\lambda ,%
\protect\beta ).$}
\label{ns-N}
\end{figure}

In Fig.~\ref{ns-N}, we show the dependence of the spectral index on the
number of e-folds of inflation, for the same range of values of the
parameters $(\lambda ,\beta )$ of Fig.~\ref{r-ns}. We can observe that there
is a range of values of $n_{s}$ and $r$ that is compatible with the WMAP3
analysis. For small numbers of e-folds of inflation, compatibility is
assured for large values of the parameter $\lambda $. 

\begin{figure}[tbp]
\includegraphics[width = 8.5cm]{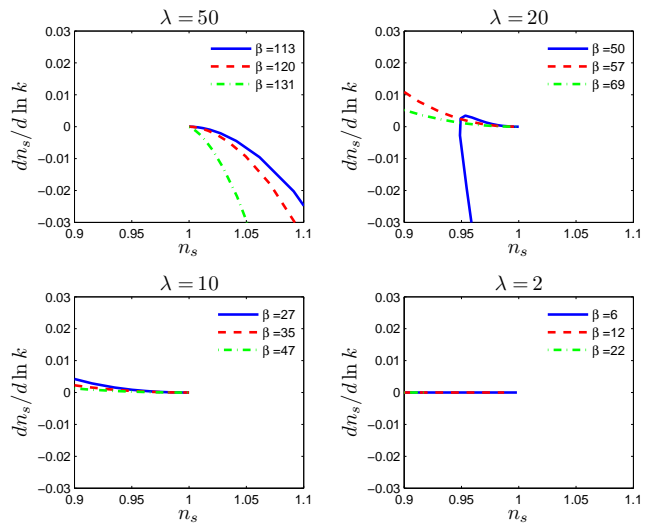}
\caption{{Trajectories in the $n_{s}-(dn_{s}/d\ln k)$ plane, for different
combinations of the parameters $(\protect\lambda ,\protect\beta )$.}}
\label{dnsdlnk-ns}
\end{figure}

The running of the spectral index, to lowest order in slow-roll, is given by 
\begin{equation}
\frac{dn_{s}}{d\ln k}=-8\epsilon ^{2}+10\epsilon \eta -2\xi ^{2}\,.
\end{equation}%
From Fig.~\ref{dnsdlnk-ns}, we observe that for certain combinations of the
parameters $(\lambda ,\beta ),$ the running of the spectral index can be
negative, which is favored by WMAP3. 

\subsection{Multi-field inflation}

In the original model of assisted power-law inflation, with combinations of
pure exponential potentials, the dynamics can be interpreted by performing a
rotation in the field space decomposing the fields into a weighted mean
field that sets the direction of motion, and a set of fields orthogonal to
it. The potential for the orthogonal directions has a minimum, therefore
assuring the stability of the scaling solution obtained in these models. We
expect the same to be true in the case of logamediate inflation, although a
scaling solution is not attained (i.e. the ratio between the potential and
kinetic energies is\textit{\ not} a constant throughout the evolution)
because it is reasonable to assume that the ratios of the kinetic energies
of the fields can reach a constant value as these potentials encounter a
valley along each defined direction. Moreover, by increasing the number of
fields we are increasing the Hubble damping in the equations of motion for
these fields, hence preventing them from running away. Consequently, slow
roll evolution is reinforced. We will see now that the combined evolution of
an ensemble of fields can be mimicked by one field, as in the previous
section, where the parameter $A$ depends on the steepness on the potential
in all the original field-space directions.

We will now generalize the previous logamediate inflationary solution for
the scalar potential to include an arbitrary number of scalar fields. To
this end, and motivated by the single-field case, we will assume that at
late times we can express the cosmic time as a weighted sum of $m_{q}$
different scalar fields, $\phi _{i}$, in the form 
\begin{equation}
\ln t=\sum_{i}^{m_{q}}\alpha _{qi}\,\phi _{i}^{2/(\lambda +1)}\,,
\label{lnt}
\end{equation}%
so that when the potential is written in terms of time, $t$, it becomes 
\begin{equation}
V=\sum_{q}k_{q}\,(\ln t)^{2(\lambda -1)}\frac{1}{t^{2}}\,,
\end{equation}%
where the $k_{q}$ are constants that depend on the parameters $\alpha _{qi}$
and will be determined in what follows. Hence, we will assume that the
general form of the scalar potential that has this behavior at late times
can be written in terms of the fields as
\begin{eqnarray}
\label{genpot}
V &=&\sum_{q}^{n}v_{q}\left[ \sum_{i}^{m_{q}}\alpha _{qi}\,\phi
_{i}^{2/(\lambda +1)}\right] ^{2(\lambda -1)} \nonumber \\
&~& \times \exp \left[ -2\sum_{i}^{m_{q}}\alpha _{qi}\,\phi _{i}^{2/(\lambda +1)}%
\right] \,,
\end{eqnarray}%
where the $v_{q}$ are arbitrary constants.

We are mainly interested in those situations where all fields play an
important role in the evolution. More specifically, we will focus on models
for which the ratio of the kinetic energies of the fields reaches a constant
value at late times. Since we must have

\be
2\dot{H}=-\sum_{i}^{m}\dot{\phi}_{i}^{2}=2A\lambda (\lambda -1)(\ln
t)^{\lambda -2}t^{-2},
\ee
where $m$ is the total number of fields, we search for solutions where 
\begin{equation}
\dot{\phi}_{i}=c_{i}\left( \ln t\right) ^{(\lambda -1)/2}\frac{1}{t}\,,
\label{dotphii}
\end{equation}%
such that 
\begin{equation}
\sum_{i}^{m}c_{i}^{2}=2A\lambda \,.  \label{sumci}
\end{equation}%
The late-time solutions for the $\phi _{i}$ are obtained by integrating Eq.~(%
\ref{dotphii}), so 
\begin{equation}
\phi _{i}=\frac{2c_{i}}{\lambda +1}\left( \ln t\right) ^{(\lambda +1)/2}\,.
\label{phii}
\end{equation}

We can derive an additional condition for the $c_{i}$ coefficients by
combining Eqs.~(\ref{lnt}) and (\ref{phii}): 
\begin{equation}
\sum_{i}^{m_{q}}\alpha _{qi}\,c_{i}^{2/(\lambda +1)}=\left( \frac{\lambda +1%
}{2}\right) ^{2/(\lambda +1)}\,.  \label{addrelation}
\end{equation}%
We then go on to compute the various $c_{i}$ and $k_{q}$ in terms of the
parameters $\alpha _{qi}$ in the potential. Substituting (\ref{phii}) in the
equations of motion for each of the fields, which satisfy 
\begin{equation}
\ddot{\phi}_{i}+3H\dot{\phi}_{i}+\frac{\partial V}{\partial \phi _{i}}=0\,,
\end{equation}%
we obtain the set of relations 
\begin{equation}
3A\lambda ^{2}c_{i}=2\left( \frac{2}{\lambda +1}\right) ^{2/(\lambda
+1)}\sum_{q}^{n}k_{q}\,\alpha _{qi}\,c_{i}^{-\frac{\lambda -1}{\lambda +1}%
}\,.  \label{cieq}
\end{equation}%
Multiplying by $c_{i}$, and using constraints (\ref{sumci}) and (\ref%
{addrelation}) we obtain

\[
\sum_{q}^{n}k_{q}=3A^{2}\lambda ^{2}, 
\]%
as expected, by comparing Eqs.~(\ref{v_t}) and (\ref{genpot}).

Equation (\ref{cieq}) can also be rewritten as 
\begin{equation}
c_{i}^{2/(\lambda +1)}=\left[ \frac{2}{3A\lambda }\left( \frac{2}{\lambda +1}%
\right) ^{2/(\lambda +1)}\right] ^{1/\lambda }\left(
\sum_{q}^{n}k_{q}\,\alpha _{qi}\right) ^{1/\lambda }\,.  \label{cieq2}
\end{equation}%
Multiplying by $\alpha _{ri}$, summing over all fields, and using relation (%
\ref{addrelation}), we obtain a set of constraints that must be satisfied by
the various scales of the potential $k_{q}$: 
\begin{equation}
\sum_{i}^{m}\left( \sum_{q}^{n}k_{q}\,\alpha _{qi}\right) ^{1/\lambda
}\alpha _{ri}=\left( \frac{\lambda +1}{2}\right) ^{2/\lambda }\left( \frac{%
3A\lambda }{2}\right) ^{1/\lambda }\,.  \label{system}
\end{equation}%
In general, it is a difficult task to compute these quantities, but in the
case where the fields only appear in one of the terms in the potential, for
a given $\phi _{i},$ all the $\alpha _{qi}$ vanish except one. In this case,
Eqs.~(\ref{system}) can be simplified to give 
\begin{equation}
k_{q}=\left( \frac{\lambda +1}{2}\right) ^{2}\left( \frac{3A\lambda }{2}%
\right) \left( \sum_{i}^{m_{q}}\alpha _{qi}^{(\lambda +1)/\lambda }\right)
^{-\lambda }\,.
\end{equation}%
We have already seen that $\sum_{q}^{n}k_{q}=3A^{2}\lambda ^{2}$; hence, we
can write the parameter $A$ in terms of the coefficients, $\alpha _{qi}$, in
the potential (\ref{genpot}): 
\begin{equation}
A=\frac{1}{2\lambda }\left( \frac{\lambda +1}{2}\right)
^{2}\sum_{q}^{n}\left( \sum_{i}^{m_{q}}\alpha _{qi}^{(\lambda +1)/\lambda
}\right) ^{-\lambda }\,.  \label{lateA1}
\end{equation}%
From this expression we can see that the value of $A$ increases if we
increase the number of terms in the potential and decreases if we increase
the number of scalar fields in each of the terms. This behavior is very
similar to that encountered in assisted power-law inflation driven by a
combination of pure exponential potentials \cite{Copeland:1999cs}.

From Eq.~(\ref{cieq2}), we are now ready to compute the $c_i$: 
\begin{equation}
c_i^{2\lambda/(\lambda+1)} = \left(\frac{\lambda+1}{2}\right)^{2\lambda/(%
\lambda+1)} \alpha_{qi} \left(\sum_j^{m_q}
\alpha_{qj}^{(\lambda+1)/\lambda}\right)^{-\lambda} \,.
\end{equation}

\begin{figure}[tbh]
\includegraphics[width = 8.5cm]{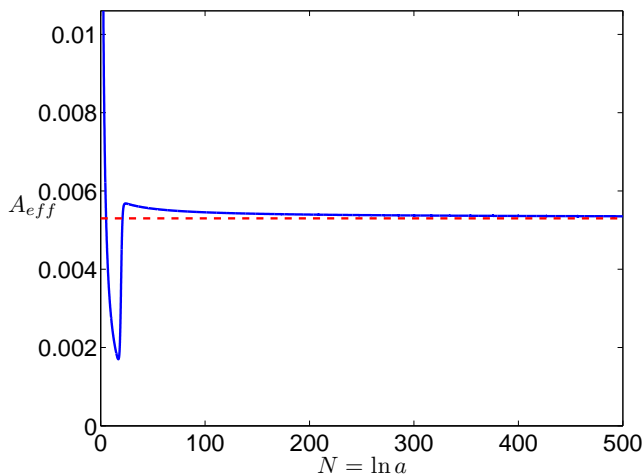}
\caption{Evolution of $A_{\mathrm{eff}}$ (solid line) compared with the
late-time value expected from Eq.~(\protect\ref{lateA1}) (dashed line). The
actual potential used is $V=\sum_{q}v_{q}(\sum_{i}\protect\alpha _{qi}%
\protect\phi _{i}^{2/(\protect\lambda +1)})^{2(\protect\lambda -1)}\exp
[-2\sum_{i}\protect\alpha _{qi}\protect\phi _{i}^{2/(\protect\lambda +1)}]$
with $\protect\alpha _{11}=3.6$, $\protect\alpha _{12}=3.9$, $\protect\alpha %
_{23}=3.3$, $\protect\alpha _{24}=4.2$, $\protect\lambda =2$ and $%
v_{1}=v_{2}=1$.}
\label{figAeff}
\end{figure}

In the Fig.~\ref{figAeff}, we compare the effective value of $A$ 
\begin{equation}
A_{\mathrm{eff}}=\left( \frac{-H^{2}}{\dot{H}}\right) ^{\lambda }\,\frac{1}{%
\lambda ^{\lambda }N^{\lambda -1}}\,,
\end{equation}%
determined from the numerical integration of the equations of motion, with
the value expected at late times using Eq.~(\ref{lateA1}).

\section{Intermediate inflation: $a=\exp (At^{f})$}

Intermediate inflation is defined by a scale-factor evolution of the form $%
a=\exp (t^{f})$, for which we have $H=Aft^{f-1}$. Hence, for an expanding
universe, $Af>0$. From the time derivative of the Hubble rate $\dot{H}%
=Af(f-1)t^{f-2},$ which must be negative, we conclude in addition that $f<1$%
. Consequently, we impose $A>0$ and $0<f<1$.

\subsection{Single-field inflation}

For a single scalar field, $\dot{H}=-\dot{\phi}^{2}/2$ and we have 
\begin{equation}
\dot{\phi}=\left( 2Af(1-f)\right) ^{1/2}t^{f/2-1}\,,
\end{equation}%
which gives the evolution of the field as 
\begin{equation}
\phi =\left( 8A\,\frac{1-f}{f}\right) ^{1/2}t^{f/2}\,,
\end{equation}%
and results in a time-dependence of the potential given by 
\begin{equation}
V=3\left( Aft^{f-1}\right) ^{2}-Af(1-f)t^{f-2}\,.
\end{equation}%
Substituting time for the value of $\phi $, we obtain at late times that 
\begin{equation}
V(\phi )=48\frac{A^{2}(2A\beta )^{\beta /2}}{(\beta +4)^{2}}\,\phi ^{-\beta
}\,,
\end{equation}%
where $\beta =4(1-f)/f$.

The relation between the spectral index and the ratio of scalar to tensor
perturbations takes the form 
\begin{equation}
n_{s}=1-\frac{\beta -2}{8\beta }r\,,
\end{equation}%
and it was found that this model is in agreement with the latest WMAP data 
\cite{Barrow:2006dh}.

We will now determine the effective value of $A$ when several fields are
evolving.

\subsection{Multi-field inflation}

As before, we can admit that the cosmic time can be written in terms of a
combination of the fields such that 
\begin{equation}
t=\sum_{i}^{m_{q}}\alpha _{qi}\phi _{i}^{2/f}\,,  \label{tempo}
\end{equation}%
and the potential can be written as 
\begin{equation}
V=\sum_{q}^{n}v_{q}\left( \sum_{i}^{m_{q}}\alpha _{qi}\phi _{i}^{2/f}\right)
^{2(f-1)}\,.
\end{equation}%
We must note, however, that $v_{q}$ is dependent of $\alpha _{qi}$ and not
free as in the previous model. Indeed, the potential can be rewritten in
terms of the parameters $b _{qi}=v_{q}^{1/(2f-2)}\alpha _{qi} $ in the form 
\begin{equation}
V=\sum_{q}^{n}\left( \sum_{i}^{m_{q}}b_{qi}\,\phi _{i}^{2/f}\right)
^{2(f-1)}\,.
\end{equation}

Scalar potentials of this form, however, have a ridge in field space rather
than a valley as in the previous example, consequently, any solution for
which the ratio of the kinetic energies of the fields is a constant, is
unstable. For this reason we will focus on the simplified class of
potentials given by 
\begin{equation}
V = \sum_i^m \alpha_i \, \phi_i^{4(f-1)/f} \,.
\end{equation}
Admitting that the system attains a regime where all the fields are
important in the evolution of the universe, we write 
\begin{equation}
\phi_i =\frac{2}{f}c_{i}t^{f/2}\,,  \label{phiib}
\end{equation}%
and by requiring 
\begin{equation}
2\dot{H}=-\sum_{i}^{m}\dot{\phi}_{i}^{2}=2Af(1-f)t^{f-2} \,,
\end{equation}
we have the condition 
\begin{equation}
\sum_{i}^{m}c_{i}^{2}=2Af(1-f)\,.  \label{sumcib}
\end{equation}

Substituting Eq.~(\ref{phiib}) in the equations of motion results in the
following set of relations; 
\begin{equation}
3Afc_{i}=4\frac{1-f}{f}\left( \frac{2}{f}\right) ^{3-4/f}\alpha
_{i}\,c_{i}^{3-4/f}\,.  \label{cieqb}
\end{equation}%
Multiplying by $c_{i}$ and using condition (\ref{sumcib}) we obtain 
\begin{equation}
3A^{2}f^{2}=\left( \frac{2}{f}\right) ^{4-4/f}\sum_{i}^{m}\alpha
_{i}\,c_{i}^{4(f-1)/f}\,,  \label{3a2f2}
\end{equation}%
Using Eq.~(\ref{cieqb}) to write $c_{i}$ in terms of $\alpha _{i}$ and
substituting into Eq.~(\ref{3a2f2}) we have that 
\begin{equation}
A=\frac{1}{3^{f/2}f}\left( \frac{f^{2}}{8(1-f)}\right) ^{1-f}\left(
\sum_{i}^{m}\alpha _{i}^{f/(2-f)}\right) ^{(2-f)/2}\,,  \label{Ab}
\end{equation}%
We see that also in intermediate inflation, the effective value of $A$
increases by increasing the number of fields. Upon substitution back into
Eq.~(\ref{cieqb}), the coefficients $c_{i}$ are given by 
\begin{equation}
c_{i}^{2-4/f}=\frac{3^{(2-f)/2}}{f\alpha _{i}}\left( \frac{2}{f}\right)
^{f-4+4/f}\left( 4\frac{f-1}{f}\right) ^{f-2}\,.
\end{equation}

For a specific model, we can now compare the late-time numerical evolution
of the effective $A_{\mathrm{eff}}$ 
\begin{equation}
A_{\mathrm{eff}}=\left( \frac{-\dot{H}}{H}\right) ^{f}\,\frac{N}{(1-f)^{f}}%
\,.
\end{equation}%
against the expected value given by Eq.~(\ref{Ab}). An exercise of this kind
would lead to an equivalent evolution to the one found in Fig.~\ref{figAeff}.

\section{Conclusions}

We have analysed a new inflationary scenario named `logamediate' inflation and revisited the
scenario of intermediate inflation. We have demonstrated that both lead to
phenomenologically viable models of inflation as there are wide regions of
parameter space compatible with the latest CMB observations. Then, in each
of these two scenarios, we generalized the solutions obtained for the scalar
potentials to allow an ensemble of scalar fields to participate in the
dynamics. As in the original model of assisted inflation for pure
exponential potentials, we found that also in intermediate and logamediate
inflation the fields behave as a group and the late-time dynamics is
dictated by the steepness of the potential in all field space directions.
However, unlike the original assisted inflation case, the cooperative
behavior does not determine if inflation is more or less probable as we
increase the number of fields. The reason for this difference is that the
parameters that affect the condition for acceleration of the universe are $%
\lambda >1$ and $f<$ $1$, which are required to be fixed to a certain value
for all fields. The only effective parameter that depends on the parameters
of the potential is $A,$ whose value, however, does not influence the
criteria for accelerated expansion to occur.

\begin{acknowledgments}
N.J.N was supported by PPARC.
\end{acknowledgments}

\end{document}